\documentclass[aps,prb,amsmath,amssymb,reprint,superscriptaddress,preprintnumbers,showpacs,intlimits,longbibliography]{revtex4-1}
\usepackage{bm,latexsym,mathrsfs,enumerate,color}
\usepackage[mathcal]{euscript}
\usepackage[breaklinks=true,unicode=true,urlcolor = blue,colorlinks = true,citecolor = blue,linkcolor = blue]{hyperref}
\usepackage{graphicx}
\usepackage{todonotes}
\usepackage{adjustbox}
\usepackage{array,ragged2e}
\renewcommand{\vec}[1]{\bm{#1}}
\usepackage[normalem]{ulem}

\bibpunct{[}{]}{,}{n}{}{}

\begin{document}
%
%
\title{Spin eigen-excitations of the magnetic skyrmion and problem of the effective mass}

\author{Volodymyr P. Kravchuk}
\email{vkravchuk@bitp.kiev.ua}
\affiliation{Bogolyubov Institute for Theoretical Physics of National Academy of Sciences of Ukraine, 03680 Kyiv, Ukraine}
\affiliation{Leibniz-Institut f{\"u}r Festk{\"o}rper- und Werkstoffforschung, IFW Dresden, D-01171 Dresden, Germany}


\author{Denis D. Sheka}
\affiliation{Taras Shevchenko National University of Kyiv, 01601 Kyiv, Ukraine}


\author{Ulrich K. R{\"o}{\ss}ler}
\affiliation{Leibniz-Institut f{\"u}r Festk{\"o}rper- und Werkstoffforschung, IFW Dresden, D-01171 Dresden, Germany}

\author{Jeroen~van~den~Brink}
\affiliation{Leibniz-Institut f{\"u}r Festk{\"o}rper- und Werkstoffforschung, IFW Dresden, D-01171 Dresden, Germany}
\affiliation{Institute for Theoretical Physics, TU Dresden, 01069 Dresden, Germany}

%
%

\author{Yuri Gaididei}
\affiliation{Bogolyubov Institute for Theoretical Physics of National Academy of Sciences of Ukraine, 03680 Kyiv, Ukraine}

\date{\today}

%
%
%
%
\begin{abstract}
Properties of magnon modes localized on a ferromagnetic skyrmion are studied. Three types of possible asymptotic behavior of the modes eigenfrequencies are found for the case of large skyrmion radius $R_s$, namely $\omega_0\sim R_s^{-2}$ for the breathing mode, $\omega_{-|\mu|}\sim R_s^{-1}$ and $\omega_{|\mu|}\sim R_s^{-3}$ for modes with negative and positive azimuthal quantum numbers, respectively. A number of properties of the magnon eigenfunctions are determined. This enables us to demonstrate that the skyrmion dynamics based on the traveling wave model is described by the massless Thiele equation.
\end{abstract}

\maketitle

\section{Introduction}

Chiral magnetic skyrmion \cite{Liu16a,Seidel16,Wiesendanger16,Nagaosa13,Bogdanov89,Bogdanov94,Bogdanov01,Ivanov90a} are particle-like topological solitons with an integer topological charge (skyrmion number). In contrast to magnetic vortex \cite{Kamenetskii08,Mertens00,Papanicolaou91}, which has half-integer topological charge, skyrmion is a truly localized excitation. As a consequence, the dynamics of the skyrmion is insensitive to the size and shape of a sufficiently large sample, and remote skyrmions do not interact with each other, except the weak stray field effects. This is in strong contrast to magnetic vortices. The possibility to manipulate skyrmions as an ``individual particles'' together with the topological stability of this ``particles'' results in a large number of studies during the last years, where skyrmions are considered as key elements for nonvolatile memory and logic devices \cite{Fert13,Sampaio13,Zhang15,Krause16,Kang16,Wiesendanger16}.
This composed the new area of spintronics -- skyrmionics \cite{Krause16}.

On the other hand, magnon spintronics \cite{Chumak15} is another recently emergent area of spintronics, which is intensively developing in parallel to the skyrmioncs. Magnon transistor \cite{Chumak14} and a number of the wave-based logic devices \cite{Chumak15,Schneider08} are proposed as key elements for the concept of the wave-based computing \cite{Chumak15}. In this regard, the combination of these two trends (e.g. skyrmion-based nonvolatile memory and magnod-baged logic devices) can give a significant push in the development of magnetic computer elements alternatively to the existed semiconductor chips. However, in this case, a deep understanding of properties of the skyrmion magnon modes is required.
 
The spectrum of linear excitations of any system reflects its fundamental properties, e.g. it enables one to find out conditions of instabilities and analyze their types. Knowledge of spectral dependencies opens wide opportunities for the resonance experimental techniques \cite{Klein08,Castel12,Novosad05} The presence of the skyrmion resonances can be considered as a proof of presence of skyrmions in the system and values of the eigenfrequencies can be used for determination of values of the material parameters. 

Due to the fundamental interest and high importance for the applications the  number of studies of magnon spectra of skyrmions were recently carried out \cite{Schuette14,Lin14a,Iwasaki14a,Schroeter15,Garst17a,Makhfudz12,Buijnsters14,Zhang17a,Mruczkiewicz17,Guslienko17}. However, most of these studies are numerical and the existing theories are far from the completeness. Some important questions are still not clear, e.g. the presence of the high-frequency gyrotropic mode, which is a counterpart to the zero translational mode. It has the same azimuthal symmetry but nonzero eigenfrequency. In some models this mode is responsible for the formal appearance of mass of the topological soliton \cite{Volkel94,Mertens97,Wysin96a,Ivanov98,Mertens00,Sheka01,Ivanov05b,Sheka06,Makhfudz12,Buijnsters14,Buettner15}, see discussion in Sec.~\ref{sec:col-var}. In most of the mentioned studies this mode  was not found \cite{Schuette14,Lin14a,Iwasaki14a,Schroeter15}. On the other hand, it was found for some cases of large-radius skyrmions \cite{Makhfudz12,Buijnsters14,Zhang17a}.

In this paper we present a detailed study and classification of the skyrmion magnon spectrum. The paper is build as follows. Section~\ref{sec:model} is declaratory in nature, here we explain our model and introduce a system of the dimensional units used in the paper, see Table~\ref{tbl:units}. In Sec.~\ref{sec:breath} we consider the simplest case of the radially symmetrical magnon mode, so called breathing mode. The general study of the magnon spectrum is presented in Sec.~\ref{sec:magnons}. Here we obtain the asymptotic behavior of the eigenfrequencies  of all localized modes. Additionally we find out a number of properties of the modes eigenfunctions, e.g. the orthogonality condition. These properties are utilized in Sec.~\ref{sec:col-var}, where we use the collective variables approach in order to demonstrate that the skyrmion dynamics based on the traveling wave model is described by the \textit{massless} Thiele equation.


\section{Model and static equilibrium solutions}\label{sec:model}
\begin{table}
	\begin{tabular}{>{\RaggedRight\arraybackslash}p{2.5cm}>{\RaggedRight\arraybackslash}p{2.8cm}>{\RaggedRight\arraybackslash}p{2.7cm}}\hline\hline
		quantity & unit of \hfill \break measurement & value for Pt/Co/AlO$_x$ layer structure~\cite{Rohart13} \\ \hline
		length & $\ell=\sqrt{A/K}$ & 5.6~nm \\
		time & $\Omega_0^{-1}=\left(\frac{2K\gamma_0}{M_s}\right)^{-1}$ & $98\,\text{ps}$ \hfill \break ($\Omega_0=10.2$~GHz) \\[3pt]
		energy & $E_\textsc{bp}=8\pi A L$ & {\raggedleft
		$4\times10^{-19}\,\text{J}$ } \hfill \break (for $L=1$~nm)\\
		mass & $\mathcal{M}^\star=L\frac{M_s^2}{K\gamma_0^2}$ & $2.4\times10^{-21}$~kg \hfill \break (for $L=1$~nm)\\
		DMI strength & $\sigma=\sqrt{AK}$ & 2.9~mJ/m$^2$ \\
		\hline\hline
	\end{tabular}	
	\caption{Units of physical quantities, which are used in this paper. Values are calculated for the following material parameters: $A=1.6\times10^{-11}$~J/m, $M_s=1.1\times10^6$~A/m and $K=K_0-2\pi M_s^2=5.1\times10^5$~J/m$^3$ with $K_0=1.3\times10^6$~J/m$^3$ being the intrinsic magneto-crystalline anisotropy.}\label{tbl:units}
\end{table}

Here we consider the case of chiral skyrmion, which is stabilized in perpendicular easy-axial ferromagnetic film without external magnetic field influence. The advantage of the field-free case is that the skyrmion can have an arbitrary large radius. However, the problem of the magnon spectrum is still not studied systematically for this case. 

In our model we take into account three contributions into the total energy of the ferromagnetic film 
\begin{equation}\label{eq:E}
E=L\int\left[A \mathscr{E}_\text{ex}+K(1-m_z^2)+D\mathscr{E}_\textsc{d}\right]\mathrm{d}S,
\end{equation}
here the integration is performed over the film area and $L$ is the film thickness, which is assumed to be small enough to ensure uniformity of the magnetization in the perpendicular direction. The first term of the integrand is the exchange energy density with $\mathscr{E}_\text{ex}=\sum_{i=x,y,z}(\partial_i\vec{m})^2$, and $A$ is the exchange constant. Here $\vec{m}=\vec{M}/M_s$ is the unit magnetization vector with $M_s$ being the saturation magnetization. The second term is the perpendicular easy-axis anisotropy with $K>0$ and $m_z=\vec{m}\cdot\hat{\vec{z}}$ is the magnetization component normal to the surface. The last term represents the Dzyaloshinskii-Moriya interaction (DMI) with $\mathscr{E}_\textsc{d}=m_z\vec\nabla\cdot\vec{m}-\vec{m}\cdot\vec{\nabla}m_z$. This inhomogeneous DMI is associated with the $C_{\infty v}$ symmetry of (idealized) ultrathin films \cite{Crepieux98,Bogdanov01,Thiaville12} or bilayers \cite{Yang15}, stems from the relativistic spin-orbit coupling \cite{Dzyaloshinsky57,Dzyaloshinski64}, and selects a fixed sense of rotation for any twisted noncollinear magnetization configuration.  

In the following we introduce a system of dimensionless units, explained in the Table~\ref{tbl:units}. The proposed units of measurement have the following physical sense: magnetic length $\ell$ is a typical width of a domain wall in the given system with vanishing DMI and $\sigma$ is the energy per unit area of such a domain wall, $\Omega_0$ is frequency of the uniform ferromagnetic resonance with $\gamma_0$ being the gyromagnetic ratio, $E_\textsc{bp}$ is energy of Belavin-Polyakov soliton \cite{Belavin75}, which coincides with energy of a skyrmion with infinitesimally small radius.


The constraint $|\vec{m}|=1$ is encoded by using the spherical angular parameterization $\vec{m}=\sin\theta\cos\phi\hat{\vec{x}}+\sin\theta\sin\phi\hat{\vec{y}}+\cos\theta\hat{\vec{z}}$.
Magnetization dynamics is described by Landau-Lifshitz equations
\begin{equation}\label{eq:LL}
\sin\theta\dot\phi=4\pi\frac{\delta \mathcal{E}}{\delta\theta},\qquad \sin\theta\dot\theta=-4\pi\frac{\delta\mathcal{E}}{\delta\phi},
\end{equation}
where the overdot denotes the derivative with respect to the dimensionless time $\tau=t\Omega_0$, and $\mathcal{E}=E/E_\textsc{bp}$ is the dimensionless energy. Here and below, all distances are measured in units $\ell$, in accordance with Table~\ref{tbl:units}. 

For the proposed system of units the dimensionless DMI constant $d=D/\sigma$ is the only parameter, which controls the system. The described model is characterized by the critical value of the DMI constant $d_0=4/\pi$, which separates two ground states, namely the uniform state $\vec{m}=\pm\hat{\vec{z}}$ for the case $|d|<d_0$, and helical periodical state for $|d|>d_0$  \cite{Dzyaloshinski64,Dzyaloshinski65,Bogdanov94,Rohart13,Komineas15c,Leonov16}.
For the case $0<|d|<d_0$  Eqs.~\eqref{eq:LL} have a stationary skyrmion solution  $\theta=\Theta(\rho)$ and $\phi=\Phi(\chi)$, where $\Phi=\chi+\Psi_0$ and $\{\rho,\chi\}$ are polar coordinates. This is an excitation of the uniform ground state. The skyrmion profile is determined by equation \cite{Rohart13,Komineas15c,Leonov16,Bogdanov94,Bogdanov89} 
\begin{equation}\label{eq:Theta}
\begin{split}
&\Delta_\rho\Theta-\sin\Theta\cos\Theta\left(1+\frac{1}{\rho^2}\right)+\frac{|d|}{\rho}\sin^2\Theta=0,\\
&\Theta(0)=\pi,\qquad\Theta(\infty)=0.
\end{split}
\end{equation}
where $\Delta_\rho f=\rho^{-1}\partial_\rho(\rho\,\partial_\rho f)$ is radial part of the Laplace operator. For the case of boundary conditions \eqref{eq:Theta} one has $\Psi_0=0$ if $d>0$, and $\Psi_0=\pi$ if $d<0$. In the other words, the considered type of DMI induces so called N\'{e}el (hedgehog) skyrmion.
The case of an alternative boundary conditions $\Theta(0)=0$ and $\Theta(\infty)=\pi$ is not considered here, because it does not result in fundamentally new properties of the magnon spectra compared to \eqref{eq:Theta}.

\section{Large-radius skyrmion and radially symmetrical mode}\label{sec:breath}

\begin{figure}
	\includegraphics[width=\columnwidth]{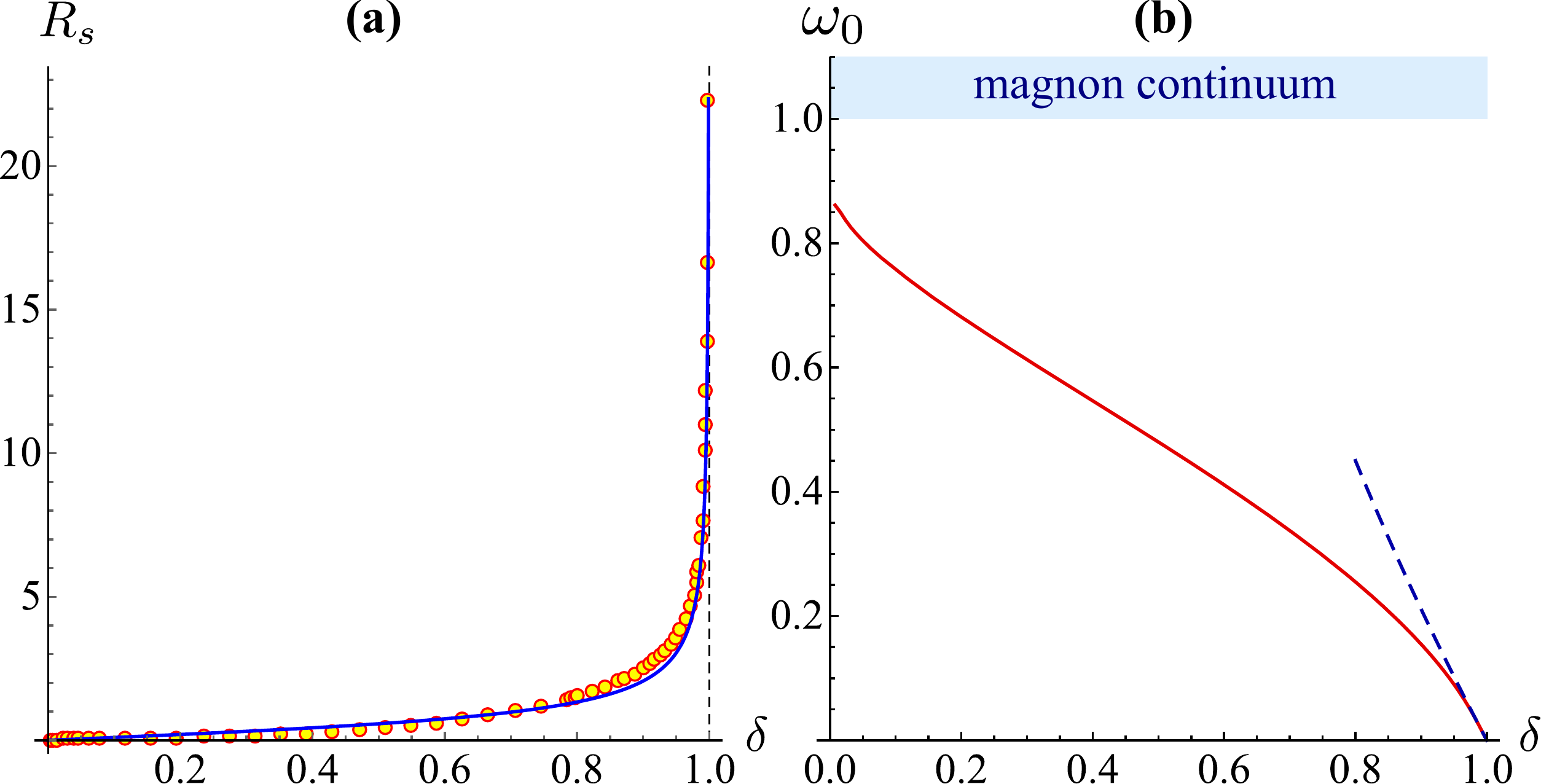}
	\caption{(a) Skyrmion radius as a function of the normalized DMI constant. Line shows the estimation \eqref{eq:Rs} while dots correspond to the exact numerical solutions of Eq.~\eqref{eq:Theta}. (b) Breathing mode eigenfrequency (solid line) and asymptotics \eqref{eq:w0-large-Rs} (dashed line) for the case of large skyrmion radius ($\delta\lessapprox1$).}\label{fig:Rs}
\end{figure}

The skyrmion profile is a localized function \cite{Rohart13,Komineas15c,Leonov16,Bogdanov94}, this enables one to introduce the skyrmion radius $R_s$ as solution of the equation $\cos\Theta(R_s)=0$. Skyrmion radius $R_s$ strongly depends on the DMI constant \cite{Rohart13,Bogdanov94}: $R_s\to0$ when $d\to0$ (skyrmion collapse), while $R_s\to\infty$ when $d\to\pm d_0$, see Fig.~\ref{fig:Rs}(a). For a large-radius skyrmion $R_s\gg1$ one can easily estimate the asymptotic behavior of $R_s$. In this case the skyrmion profile is well described by the circular domain wall Ansatz 
\begin{equation}\label{eq:Ansatz-large}
\cos\theta=\tanh\frac{\rho-R_s}{\Delta},\qquad\phi=\chi+\Psi
\end{equation}
Here the skyrmion radius $R_s$, phase $\Psi$ and width $\Delta$ are treated as collective variables. This is in contrast to the number of the previous papers \cite{Sheka01,Makhfudz12,Rohart13}, where the model \eqref{eq:Ansatz-large} was used for the constant $\Delta$. 
Normalized energy of the skyrmion, which is described by the model \eqref{eq:Ansatz-large}, reads
\begin{equation}\label{eq:energy-large}
\mathcal{E}\approx\frac12\left(\underbrace{\frac{R_s}{\Delta}+\frac{\Delta}{R_s}}_{\text{exchange}}-\underbrace{2\delta R_s\cos\Psi}_{\text{DMI}}+\underbrace{\Delta R_s}_{\text{anisotr.}}\right),
\end{equation}
where $\delta=d/d_0$ is normalized DMI constant. Minimum of the energy \eqref{eq:energy-large} is reached for the following equilibrium values of the collective variables: $\Delta_0=|\delta|$ and $\Psi_0=0$ ($\Psi_0=\pi$) when $\delta>0$ ($\delta<0$).
And the skyrmion radius reads 
\begin{equation}\label{eq:Rs}
R_s\approx|\delta|/\sqrt{1-\delta^2}.
\end{equation}
In the limit case $\delta\to1$ the estimation \eqref{eq:Rs} is transformed to the previously proposed one \cite{Rohart13} $R_s\approx1/\sqrt{2(1-\delta)}$, which was obtained for the constant width $\Delta=1$. The obtained estimation agrees well with the exact values of $R_s$ for the all range of the DMI constant, not only for the case $R_s\gg1$, see Fig.~\ref{fig:Rs}(a).

Let us now consider excitation of the radially symmetrical mode. This can be easily done by considering the time dependent dynamics of the collective variables $R_s(t)$ and $\Psi(t)$ in vicinity of their equilibrium values. 
To this end we treat the equations of motion \eqref{eq:LL} as extrema of the action functional $S=\int\mathcal{L}\text{d}\tau$ with the Lagrange function
\begin{equation}\label{eq:Lagrange}
\mathcal{L}=\frac{1}{4\pi}\int(1-\cos\theta)\dot\phi\,\mathrm{d}\mathcal{S}-\mathcal{E}.
\end{equation}
Ansatz \eqref{eq:Ansatz-large} represents a circularly closed domain wall. It is well known \cite{Clarke08,Hillebrands06,Schryer74,Thiaville02a,Malozemoff79} that dynamics of the width $\Delta$ of the 1D domain wall has essentially different timescale as compared to the timescale of the domain wall position ($R_s$) and phase ($\Psi$), namely $\Delta$ fast relaxes  towards its equilibrium value determined by the much slower variables $R_s$ and $\Psi$. Therefore, an assumption that $\Delta$ is a ``slave'' variable works wery well in the most cases. Here we assume that the same is true for the circularly closed domain wall if $R_s\gg1$. In this case the substitution of Ansatz \eqref{eq:Ansatz-large} into \eqref{eq:Lagrange} results in the following effective Lagrange function
\begin{equation}\label{eq:Leff}
\begin{split}
&\mathcal{L}^{\mathrm{eff}}=\dot\Psi\,R_s^2-\mathcal{E}^{\mathrm{eff}},\\
&\mathcal{E}^{\mathrm{eff}}={R_s}\left(\frac{1}{|\delta|}-2\delta\cos\Psi+|\delta|\right)+\frac{|\delta|}{R_s}.
\end{split}
\end{equation}
As follows from \eqref{eq:Leff} the skyrmion area is the canonically conjugated momentum for the skyrmion phase. The effective Lagrange function \eqref{eq:Leff} generates a set of two equations of motion for $\Psi$ and $R_s$, which have a static equilibrium solutions \eqref{eq:Rs} and $\Psi_0$, which is determined above. The corresponding linear dynamics in vicinity of the equilibrium state is characterized by the eigenfrequency 
\begin{equation}\label{eq:w0-large-Rs}
\tilde\omega_0=\frac{1-\delta^2}{|\delta|}=\frac{1}{R_s\sqrt{1+R_s^2}}\approx\frac{1}{R_s^2}.
\end{equation}
The corresponding asymptotics are shown in Fig.~\ref{fig:Rs}(b) and Fig.~\ref{fig:plain-spectrum}(b).

\section{General description of the magnon spectra}\label{sec:magnons}
The dispersion relation of linear excitations (magnons) of the uniform ground state is not influenced by the DMI and coincides with the common dispersion for the easy-axis magnets $\omega=1+q^2$ with $q=k\ell$ being the dimensionless wave-number. Thus, one has the continuum spectrum with $\omega\ge1$ for the frequencies above the anisotropy-induced gap. In the following we are interested in magnons over the equilibrium skyrmion state. For this purpose we introduce small excitations of the stationary solution: $\theta=\Theta+\vartheta$, $\phi=\Phi+\varphi/\sin\Theta$. Equations \eqref{eq:LL} linearized with respect to the excitations results in
\begin{subequations}\label{eq:dev}
\begin{align}
\begin{cases}
\dot\varphi=-\Delta\vartheta+U_1\vartheta+W\partial_\chi\varphi,\\ 
-\dot\vartheta=-\Delta\varphi+U_2\varphi-W\partial_\chi\vartheta.
\end{cases}
\end{align}
Here the Laplace operator has the form $\Delta=\Delta_\rho+\rho^{-2}\partial_{\chi\chi}^2$ and the potentials are as follows

\begin{align}\label{eq:UW}
U_1=&\cos2\Theta\left(1+\frac{1}{\rho^2}\right)-\frac{|d|}{\rho}\sin2\Theta,\\ \nonumber
U_2=&\cos^2\Theta\left(1+\frac{1}{\rho^2}\right)-\Theta'^2-|d|\left(\Theta'+\frac{\sin\Theta\cos\Theta}{\rho}\right),\\ \nonumber
W=&\frac{2}{\rho^2}\cos\Theta-\frac{|d|}{\rho}\sin\Theta.\nonumber
\end{align}
\end{subequations}
Here a prime denotes derivative with respect to $\rho$.  Equations \eqref{eq:dev} have a solution $\vartheta(\rho,\chi,\tau)=f(\rho)\cos(\omega\tau+\mu\chi+\eta)$, $\varphi(\rho,\chi,\tau)=g(\rho)\sin(\omega\tau+\mu\chi+\eta)$, where the azimuthal wave number $\mu\in\mathbb{Z}$ determines the nodes number ($2|\mu|$) when moving around the skyrmion center in azimuthal direction, and $\eta$ is an arbitrary phase. The eigenfrequencies $\omega$ and the corresponding eigenfunctions $f$, $g$ are determined by the following eigenvalue problem (EVP)
\begin{subequations}\label{eq:EVP}
	\begin{align}
{\mathrm{H}}\vec{\psi}=\omega\sigma_x\vec\psi.
\end{align}
for a Hermitian operator
\begin{align}\label{eq:H}
{\mathrm{H}}=\begin{pmatrix}
-\Delta_\rho+\frac{\mu^2}{\rho^2}+U_1&\mu W\\
\mu W&-\Delta_\rho+\frac{\mu^2}{\rho^2}+U_2
\end{pmatrix}.
\end{align}
\end{subequations}
Here $\vec{\psi}=(f,g)^\textsc{t}$, and $\sigma_x$ is the first Pauli matrix. The formulated EVP \eqref{eq:EVP} structurally coincides with the corresponding EVP previously formulated for the case of magnons over precessional solitons in easy-axis magnets \cite{Sheka01,Ivanov05b,Sheka06}, magnetic vortices in easy-plane magnets \cite{Ivanov98,Sheka04}, and magnetic skyrmions \cite{Iwasaki14a,Lin14a,Schuette14,Schroeter15}.  Note that there is an alternative formulation of \eqref{eq:dev} in form of the generalized Schr{\"o}dinger equation \cite{Sheka04,Ivanov05b}, see Appendix~\ref{app:Schr} for details.

The EVP \eqref{eq:EVP} is analyzed both  analytically and numerically, for details see Appendix~\ref{app:magnons}. Equations \eqref{eq:EVP} are invariant with respect to the simultaneous change of sign of three quantities: $\omega\to-\omega$, $\mu\to-\mu$ and $f\to-f$ (or $g\to-g$). This invariance results in a symmetry of the spectrum, which simplifies classification of the modes: one can  fix either sign of $\mu$ considering both signs of $\omega$ or one can fix sign of $\omega$ considering both signs of $\mu$. Following the previous studies \cite{Ivanov98,Sheka01,Sheka04,Ivanov05b,Schuette14} we choose the latter classification and consider nonnegative frequencies $\omega\ge0$. 

\begin{figure}
	\includegraphics[width=\columnwidth]{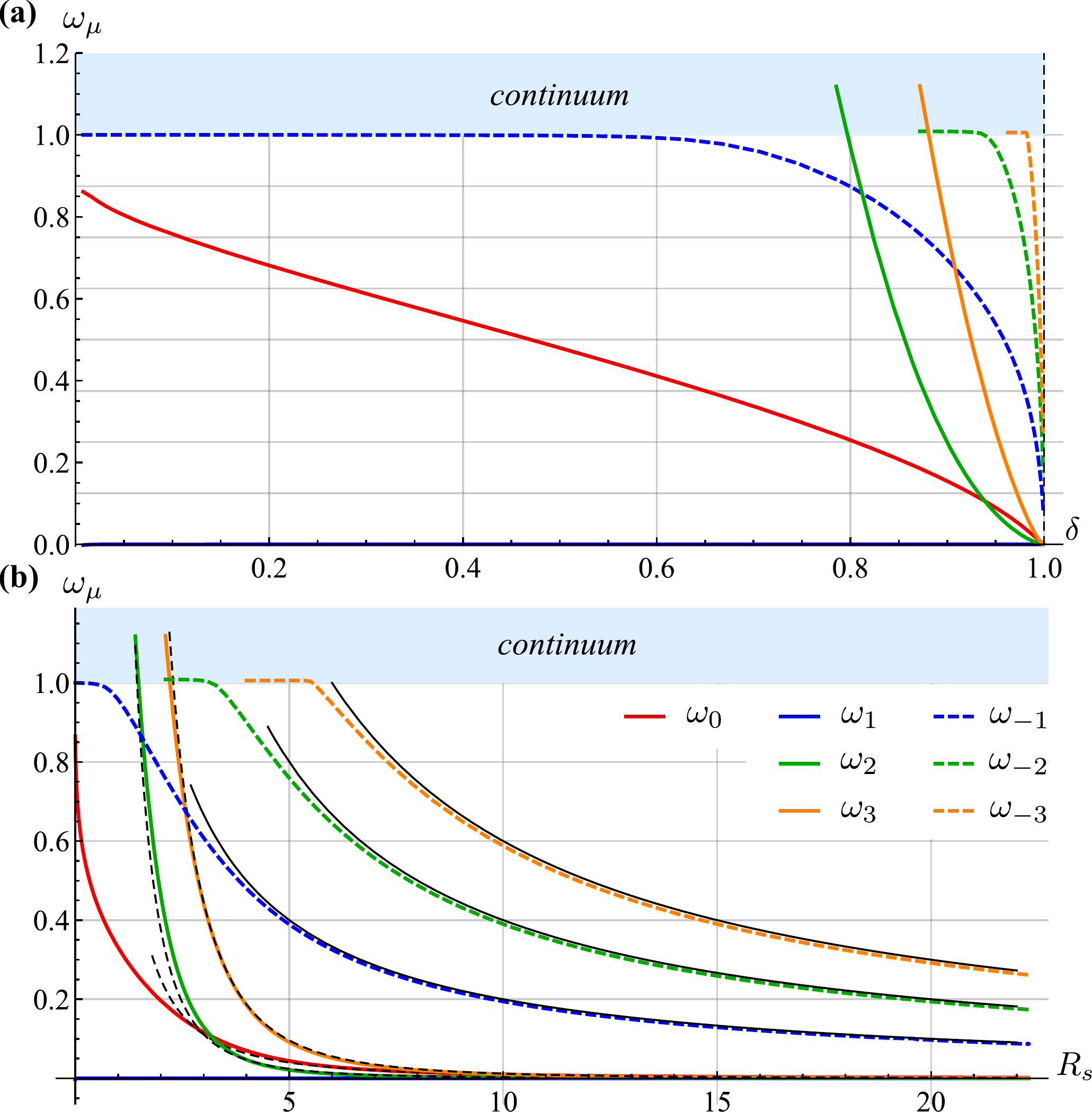}
	\caption{The eigenfrequencies $\omega_\mu$ of localized modes as functions of the normalized DMI constant $\delta=d/d_0$ (a) and skyrmion radius (b) are obtained by means of numerical solution of the EVP \eqref{eq:EVP}. Thin lines are the asymptotes \eqref{eq:asymptotics} for the case $R_s\gg1$. Modes with $|\mu|>3$ are not shown. Frequency $\omega_1$ belongs to the zero translational mode, while $\omega_{-1}$ corresponds to the high-frequency gyrotropic mode.}\label{fig:plain-spectrum}
\end{figure}

\begin{figure*}
	\includegraphics[width=0.75\textheight]{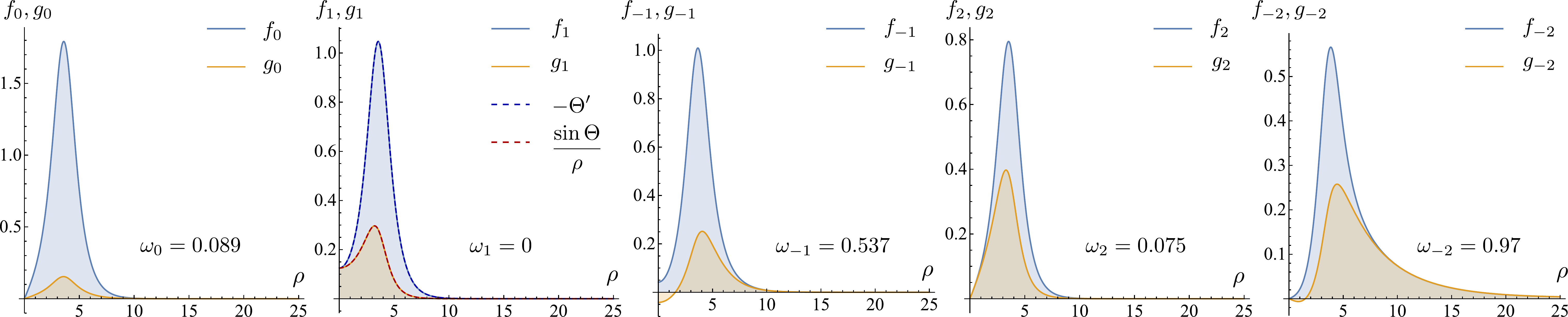}
	\caption{Eigenfunctions and eigenfrequencies obtained by means of numerical solution of \eqref{eq:EVP} for the case $\delta=0.95$ ($R_s=3.51$) and free boundary conditions. The eigenfunctions are normalized by the rule $\alpha_\mu=2$. }\label{fig:eigenfunctions}
\end{figure*}

A number of localized modes are found below the edge of the continuum spectrum. Frequencies of the localized modes strongly depend on the DMI constant, see Fig.~\ref{fig:plain-spectrum}(a); alternatively we demonstrate the dependence on skyrmion radius, Fig.~\ref{fig:plain-spectrum}(b), which can be useful for applications. For any value of the skyrmion radius (DMI constant) there are at least three localized modes: radially symmetrical (breathing) mode $|\mu=0\rangle$ and two modes $|\pm1\rangle$ which are called gyrotropic \cite{Buettner15}. Due to the translational invariance one of the gyrotropic modes has zero eigenfrequency $\omega_1=0$ (in our case this is the mode $|+1\rangle$). This mode is called translational, it has the following eigenfunctions \footnote{It can be verified by direct substitution into \eqref{eq:EVP}.} $f_1=-\Theta'$ and $g_1=\sin\Theta/\rho$, see Fig.~\ref{fig:eigenfunctions}. Although the other gyrotropic mode $|-1\rangle$, has the same azimuthal symmetry as the translational one, its eigenfunctions $f_{-1}$ and $g_{-1}$ are completely different, see Fig.~\ref{fig:eigenfunctions}. For a small radius skyrmion the eigenfrequency $\omega_{-1}$ of the mode $|-1\rangle$ practically coincides with the edge of the continuum, while for large skyrmions it is inversely proportional to the skyrmion radius, see Fig.~\ref{fig:plain-spectrum}(b). It is naturally to call the mode $|-1\rangle$ high-frequency gyrotropic mode. 

With the skyrmion radius increasing, the localized modes with higher azimuthal numbers $|\mu|\ge2$ appear in the gap. However, it is important to note that for a given $\mu$ there is no more than one localized mode. Some examples of eigenfunctions of other localized modes are shown in Fig.~\ref{fig:eigenfunctions}. Far from the skyrmion the functions $f_\mu$ and $g_\mu$ have the same asymptotic behavior because $U_1\to U_2$ when $\rho\to\infty$. Note also that the localization area of the eigenfunctions increases when the corresponding eigenfrequency reaches the bottom of the continuum.

 The eigenfunctions $f_\mu$ and $g_\mu$ has a number of properties,  which are important for the future analysis. Here we present the orthogonality condition
\begin{equation}\label{eq:norm}
\int\limits_0^\infty\rho\left[f_\mu g_{-\mu}-f_{-\mu}g_{\mu}\right]\mathrm{d}\rho=0
\end{equation}
which follows from \eqref{eq:orthog} when the symmetry between branches $\mu$ and $-\mu$ is taken into account, for details see Appendix~\ref{app:magnons}. The analogous orthogonality condition was previously obtained for  magnons over precessional solitons \cite{Sheka06}. The other properties are listed in \eqref{eq:omegas}, \eqref{eq:eps-fg}, \eqref{eq:kappa-fg}, and \eqref{eq:transl}.

The found properties of the eigenfunctions enable us to find the eigenfrequencies asymptotics for the case of a large-radius skyrmion with $R_s\gg1$: 
\begin{equation}\label{eq:asymptotics}
\omega_0\approx\frac{1}{R_s^{2}},\qquad\omega_\mu\approx\begin{cases}
2|\mu|/R_s,&\mu<0,\\
c_\mu/R_s^3,&\mu>0,
\end{cases} 
\end{equation}
where $c_\mu$ is a constant, for details see Appendix~\ref{app:magnons}. Comparison with the numerical solution enables one to suppose that $c_\mu=\mu(\mu^2-1)/2$. The asymptotics \eqref{eq:asymptotics} are shown in Fig.~\ref{fig:plain-spectrum} by the thin lines.

For the case $|\delta|\to1$ (or equivalently $R_s\to\infty$) the number of local modes grows infinitely. And in the critical point $|\delta|=1$ all the modes become unstable (the frequencies of these modes vanish).

\section{Effective mass in a collective variables approach}\label{sec:col-var}
Since the skyrmion is an exponentially localized excitation, there is an notion that its dynamics can be approximated by dynamics of a point particle with coordinates $\vec{\mathcal{R}}=\{\mathcal{X},\mathcal{Y}\}$, which are called collective coordinates. However, the way of definition of the skyrmion center $\vec{\mathcal{R}}$ is not unique. For the first time, the essential dependence of the collective coordinates dynamics on the definition of $\vec{\mathcal{R}}$ was indicated in Ref.~\onlinecite{Sheka06} for example of the precessional solitons. The recent analysis of numerically simulated skyrmion motion shows \cite{Komineas15c} that the different definitions  result in essentially different types of trajectories  $\vec{\mathcal{R}}(\tau)$. In order to describe more complicated trajectories the internal degrees of freedom of skyrmion are involved into the consideration \cite{Makhfudz12}, which results in appearance of an effective mass term in the equations of motion for the collective coordinates \cite{Makhfudz12,Buettner15}. The aim of this section is to utilize the properties of the magnon eigenfunctions in order to (i) determine the physically sound definition of the skyrmion center; (ii) clarify the appearance of the effective mass term.

Collective variable approaches \cite{Rajaraman82} are widely used for analysis of dynamics of soliton-like excitations in magnetic media \cite{Thiele73,Thiele74,Hubert98,Mertens00,Malozemoff79}. They are based on the assumption that the time dependence of the continuum magnetization vector field $\vec{m}$ can be reduced to the time dependence of a discrete set of collective variables $\vec{\xi}=\{\xi_1,\xi_2,...\}$, i.e. $\vec{m}(\vec{r},\tau)=\vec{m}(\vec{r},\vec{\xi}(\tau))$. This enables one to proceed from the couple of the partial differential equations \eqref{eq:LL} to the set of the ordinary differential equations (ODE)
\begin{equation}\label{eq:coll-var}
\begin{split}
&\sum\limits_j G_{\xi_i\xi_j}\dot{\xi_j}=\frac{\partial\mathcal{E}}{\partial\xi_i},\\
&G_{\xi_i\xi_j}=\frac{1}{4\pi}\int\sin\theta\left(\frac{\partial\theta}{\partial\xi_i}\frac{\partial\phi}{\partial\xi_j}-\frac{\partial\theta}{\partial\xi_j}\frac{\partial\phi}{\partial\xi_i}\right)\mathrm{d}\mathcal{S},
\end{split}
\end{equation}
where $\mathrm{d}\mathcal{S}=\mathrm{d}S/\ell^2$ is dimensionless area element. See also the tutorial papers Ref.~\onlinecite{Mertens00,Clarke08}. 

A case where the only variables are collective coordinates describing placement of a soliton ($\vec{\xi}=\vec{\mathcal{R}}$) is called the traveling wave model $\vec{m}(\vec{r},\tau)=\vec{m}_0(\vec{r}-\vec{\mathcal{R}}(\tau))$, where $\vec{m}_0$ is a stationary solution. Let us consider a more general ansatz
\begin{equation}\label{eq:Ansatz}
\begin{aligned}
	 &\theta^{\text{an}}(\rho,\chi,\tau)=\Theta(\rho)\\
	 &+\sum\limits_{\mu=-\infty}^{\infty}\!\!\!f_\mu(\rho)\left[\mathcal{A}_\mu(\tau)\cos\mu\chi+\mathcal{B}_\mu(\tau)\sin\mu\chi\right],\\ 
	&\phi^{\text{an}}(\rho,\chi,\tau)=\chi+\Psi_0\\ 
	&+\frac{1}{\sin\Theta(\rho)}\sum\limits_{\mu=-\infty}^{\infty}\!\!\!g_\mu(\rho)\left[\mathcal{A}_\mu(\tau)\sin\mu\chi-\mathcal{B}_\mu(\tau)\cos\mu\chi\right],
\end{aligned}
\end{equation}
which corresponds to \eqref{eq:dev-gen} but only localized modes are taken into account. Here $\Theta(\rho)$ determines the static skyrmion profile. The time-dependent collective variables $\mathcal{A}_\mu$, $\mathcal{B}_\mu$ play the role of amplitudes of the magnon modes. Thus, in our case  $\vec{\xi} = \{\mathcal{A}_k,\mathcal{B}_k\}$ with $k\in\mathbb{Z}$.

First of all, it is important to note that the ansatz \eqref{eq:Ansatz} includes the infinitesimal skyrmion displacements in sense of the traveling wave model. Indeed, the expansion of expressions
\begin{equation}\label{eq:TW}
\begin{split}
&\theta=\Theta\left(\sqrt{(x-\mathcal{X})^2+(y-\mathcal{Y})^2}\right),\\
&\phi=\arctan\frac{y-\mathcal{Y}}{x-\mathcal{X}}+\Psi_0
\end{split}
\end{equation} 
in $\mathcal{X}$, $\mathcal{Y}$ up to the linear terms and comparison with \eqref{eq:Ansatz} results in
\begin{equation}\label{eq:XY}
\mathcal{X}=\mathcal{A}_1,\qquad \mathcal{Y}=\mathcal{B}_1,
\end{equation}
if the form of the eigenfunctions $f_1=-\Theta'(\rho)$ and $g_1=\sin\Theta(\rho)/\rho$ is taken into account.

The second important observation is that the collective coordinates, which correspond to the traveling wave model, coincide with the first moment (center of mass) of the topological density:
\begin{equation}\label{eq:R-top}
\vec{\mathcal{R}}=\frac{1}{4\pi \mathcal{Q}}\int\vec{r}\,\mathcal{J}\,\mathrm{d}\mathcal{S},
\end{equation}
where $\mathcal{J}=-\vec{m}\cdot\left[\partial_x\vec{m}\times\partial_y\vec{m}\right]$ is the topological charge density and $\mathcal{Q}=(4\pi)^{-1}\!\int\!\mathcal{J}\,\mathrm{d}\mathcal{S}$ is total topological charge of the skyrmion \footnote{For the considered boundary conditions \eqref{eq:Theta} one has $\mathcal{Q}=1$.}. The expression \eqref{eq:R-top} directly follows from \eqref{eq:Ansatz} and \eqref{eq:XY}, if the orthogonality property \eqref{eq:norm} of the eigenfunctions is applied. Definition of the skyrmion center in form \eqref{eq:R-top} was used in a number of papers \cite{Komineas15c,Papanicolaou91,Papanicolaou95,Papanicolaou96,Komineas96,Papanicolaou99}.

Let us obtain the collective variable equations \eqref{eq:coll-var} in the approximation linear with respect to $\xi_i$ and $\dot{\xi}_i$. For this purpose we substitute the Ansatz \eqref{eq:Ansatz} into \eqref{eq:coll-var} and perform the integration using the orthogonality property \eqref{eq:norm} of the eigenfunctions. Finally one obtains the following components of the gyrotensor
\begin{equation}\label{eq:Gij}
G_{\mathcal{A}_\mu\mathcal{A}_{\mu'}}=G_{\mathcal{B}_\mu\mathcal{B}_{\mu'}}=0,\qquad G_{\mathcal{A}_\mu\mathcal{B}_{\mu'}}=-\frac{\alpha_{\mu}}{2}\delta_{\mu,\mu'},
\end{equation}
where $\alpha_{\mu}=\int_0^\infty\!\rho f_\mu g_\mu\text{d}\rho$. Note that $\alpha_1=2\mathcal{Q}$. Effective energy, which corresponds to the model \eqref{eq:Ansatz} has the form (up to the second order terms in $\mathcal{A}_\mu$ and $\mathcal{B}_\mu$) 
\begin{equation}\label{eq:Eeff}
\mathcal{E}=\frac{1}{8}\sum_\mu\varepsilon_\mu(\mathcal{A}_\mu^2+\mathcal{B}_\mu^2)+\mathcal{E}_0.
\end{equation}
Here $\varepsilon_\mu=\varepsilon_\mu^f+\varepsilon_\mu^g$, see \eqref{eq:eps-fg}, and $\mathcal{E}_0$ in a part of energy which is independent on the collective variables. Expression \eqref{eq:Eeff} is a result of the straightforward integration of the energy \eqref{eq:E}, when the property \eqref{eq:kappa-fg} is utilized.

Substituting \eqref{eq:Gij} and \eqref{eq:Eeff} into \eqref{eq:coll-var} one obtains the set of equations for collective coordinates
\begin{equation}\label{eq:col-var-set}
\begin{split}
\xi_i=\mathcal{A}_1:&\qquad\dot{\mathcal{B}}_1=0,\\
\xi_i=\mathcal{B}_1:&\qquad\dot{\mathcal{A}}_1=0,\\
\xi_i=\mathcal{A}_{\mu\ne1}:&\qquad\dot{\mathcal{B}}_\mu=-\omega_{\mu}\mathcal{A}_\mu,\\
\xi_i=\mathcal{B}_{\mu\ne1}:&\qquad\dot{\mathcal{A}}_\mu=\omega_{\mu}\mathcal{B}_\mu,
\end{split}
\end{equation}
where the properties \eqref{eq:omegas} and \eqref{eq:transl} were utilized. Following the terminology of Ref.~\onlinecite{Buijnsters14} one can conclude from \eqref{eq:col-var-set} that the mode $|1\rangle$ is a \emph{special zero normal mode}, which in contrast to \emph{inertial zero normal mode} \cite{Buijnsters14} does not lead to mass generation. Introducing now a driving potential $\mathcal{E}\to\mathcal{E}+\mathcal{U}(\mathcal{X},\mathcal{Y})$, which is assumed to be small enough in order not to change the eigenfunctions significantly, and taking into account \eqref{eq:XY}, one obtains from \eqref{eq:coll-var} the well known Thiele equation for a massless particle \cite{Thiele73,Mertens00,Malozemoff79}
\begin{equation}\label{eq:thiele-massless}
\left[\vec{e}_z\times\dot{\vec{\mathcal{R}}}\right]-\partial_{\vec{\mathcal{R}}}\mathcal{U}=0.
\end{equation}

Earlier it was shown \cite{Nikiforov83} that in the absence of an external driving the translational mode of the magnetic vortex in an easy-plane ferromagnet does not exist. In other words, the zero translational mode is not inertial \cite{Buijnsters14} for this case. The same results are valid for the case of magnetic skyrmion in a ferromagnetic film with the perpendicularly oriented easy-axis, when the analysis proposed in Ref.~\onlinecite{Nikiforov83} is applied. Thus the inertial mass term is not expected in equations for collective coordinates for the case of a ferromagnetic vortex as well as for the ferromagnetic skyrmion. This is in contrast to the case of antiferromagnets \cite{Velkov16} and weak ferromagnets \cite{Zvezdin10}, where the mass therm appears in a natural way, since the second order time derivative is initially present in the equation of motion written in terms of the N\'{e}el order parameter \cite{Baryakhtar79,Andreev80,Hals11,Gomonay14}. However, the inertial mass term is often used in collective coordinates equations for ferromagnetic vortices \cite{Volkel94,Mertens97,Wysin96a,Ivanov98,Mertens00}, precessional solitons \cite{Sheka01,Ivanov89,Ivanov05b,Sheka06} and skyrmions \cite{Makhfudz12,Buettner15,Ivanov90a}. In part, this ambiguity originates from alternative methods of definitions of the soliton center. A frequently used method is based on the first  moment of the perpendicular magnetization component \cite{Komineas15c,Buettner15,Buijnsters14,Makhfudz12}
\begin{equation}\label{eq:R-mz}
\textbf{R}\equiv(\textrm{X},\textrm{Y})=\frac{\int\vec{r}(1-m_z)\mathrm{d}\mathcal{S}}{\int(1-m_z)\mathrm{d}\mathcal{S}}.
\end{equation}
Thus $\textbf{R}$ signifies the ``center of mass'' of the $m_z$ distribution. 
Using \eqref{eq:Ansatz} one obtains (in a linear approximation in amplitudes $\mathcal{A}_i$, $\mathcal{B}_i$)
\begin{equation}\label{eq:XY-wrong}
\textrm{X}=c_1\mathcal{A}_1+c_{-1}\mathcal{A}_{-1},\quad\textrm{Y}=c_1\mathcal{B}_1-c_{-1}\mathcal{B}_{-1},
\end{equation}
where $c_i=\frac12\int_0^\infty\rho^3g_1(\rho)f_{i}(\rho)\text{d}\rho/\int_0^\infty\rho(1-\cos\Theta(\rho))\mathrm{d}\rho$. In this way a formal coupling between modes $|1\rangle$ and $|-1\rangle$ is introduced into the system and, as a result, an effective mass term appears \cite{Buijnsters14,Makhfudz12}. Indeed, excluding amplitudes $\mathcal{A}_{-1}$ and $\mathcal{B}_{-1}$ from equations \eqref{eq:coll-var} and \eqref{eq:XY-wrong} one obtains a set of second order ODE for $\textrm{X}$ and $\textrm{Y}$. For the case of a radially symmetrical potential $\mathcal{U}=\omega_\textsc{g}\mathcal{R}^2/2$ they can be written in the vector form
\begin{equation}\label{eq:mass-eq}
\mathcal{M}\ddot{\textbf{R}}-\left[\vec{e}_z\times\dot{\textbf{R}}\right]+k\textbf{R}=0,
\end{equation}
where $\mathcal{M}=1/(\omega_{-1}-\omega_\textsc{g})$ and $k=\omega_{-1}\omega_\textsc{g}/(\omega_{-1}-\omega_\textsc{g})$. However, it should be emphasized that treatment of the position vector $\mathbf{R}\ne\vec{\mathcal{R}}$ as a skyrmion collective coordinate is not physically sound because $\mathbf{R}$ does not describe the skyrmion displacement in the sense of the traveling wave model. 

\section{Conclusions}
Main results of this paper are as follows: (i) we obtain asymptotic behavior of localized magnon modes over the chiral skyrmion, see \eqref{eq:asymptotics}; (ii) the high-frequency gyrotropic mode is always present in the spectrum, however its frequency practically coincides with the edge of the magnon continuum, except vicinity of the critical point $|d|=4/\pi$; (iii) using the orthogonality relation \eqref{eq:norm} for the magnon eigenfunctions we show that the collective skyrmion coordinates, which describe its motion in terms of the traveling-wave model, coincides with the first moment of the topological charge distribution; (iv) in terms of these collective coordinates the  skyrmion dynamics is massless.

\section{Acknowledgments}
V.P.K. and D.D.S. acknowledge the Alexander von Humboldt Foundation for the support. This work has been supported by the DFG via SFB 1143.

\appendix

\section{Properties of the magnon spectrum}\label{app:magnons}

For each given value of $\mu$ the EVP \eqref{eq:EVP} generates a ``$\nu$-spectrum'', which is described by the eigenfunctions $f_{\mu,\nu}$ and $g_{\mu,\nu}$ with the corresponding eigenfrequencies $\omega_{\mu,\nu}$. They are determined by the EVP
\begin{equation}\label{eq:EVP-mu-nu}
{\mathcal{H}}_{\mu}\vec{\psi}_\mu=\omega_\mu\vec\psi_\mu,
\end{equation}
where $\vec{\psi}_\mu=(f_\mu,g_\mu)^\textsc{t}$ and ${\mathcal{H}}_{\mu}={\sigma}_x{\mathrm{H}}$, see \eqref{eq:H}. Operator ${\mathcal{H}}_{\mu}$ is hermitian in a Hilbert space $\mathbb{H}^\mu$ with scalar product
\begin{equation}\label{eq:scalar-product}
\langle\vec\psi_{\mu,\nu}|\vec\psi_{\mu,\nu'}\rangle\equiv\int\limits_0^\infty\rho\,\vec\psi^{\textsc{t}}_{\mu,\nu}{\sigma}_x\vec\psi_{\mu,\nu'}\,\mathrm{d}\rho.
\end{equation}
This results in real valued eigenfrequencies $\omega_{\mu,\nu}$, and \eqref{eq:scalar-product} enables us to formulate the orthogonality condition (for the given $\mu$):
\begin{equation}\label{eq:orthog}
\int\limits_0^\infty\rho\left[f_{\mu,\nu}g_{\mu,\nu'}+f_{\mu,\nu'}g_{\mu,\nu}\right]\mathrm{d}\rho=C\,\delta_{\nu,\nu'}
\end{equation}
with $C$ being the normalization constant.

Assuming that for all $\mu$ the EVP \eqref{eq:EVP-mu-nu} has no degenerate eigenvalues, one can present the general solution of \eqref{eq:dev} in form of partial wave expansion
\begin{equation}\label{eq:dev-gen}
	\begin{split}
		\vartheta=\sum\limits_{\mu=-\infty}^{\infty}\sum\limits_\nu f_{\mu,\nu}(\rho)\bigl[&\bar{A}_{\mu,\nu}\cos(\mu\chi+\omega_{\mu,\nu}\tau)\\
		&-\bar{B}_{\mu,\nu}\sin(\mu\chi+\omega_{\mu,\nu}\tau)\bigr],\\
		\varphi=\sum\limits_{\mu=-\infty}^{\infty}\sum\limits_\nu g_{\mu,\nu}(\rho)\bigl[&\bar{A}_{\mu,\nu}\sin(\mu\chi+\omega_{\mu,\nu}\tau)\\
		&+\bar{B}_{\mu,\nu}\cos(\mu\chi+\omega_{\mu,\nu}\tau)\bigr].
	\end{split}
\end{equation}

\begin{figure*}
	\includegraphics[width=0.7\textheight]{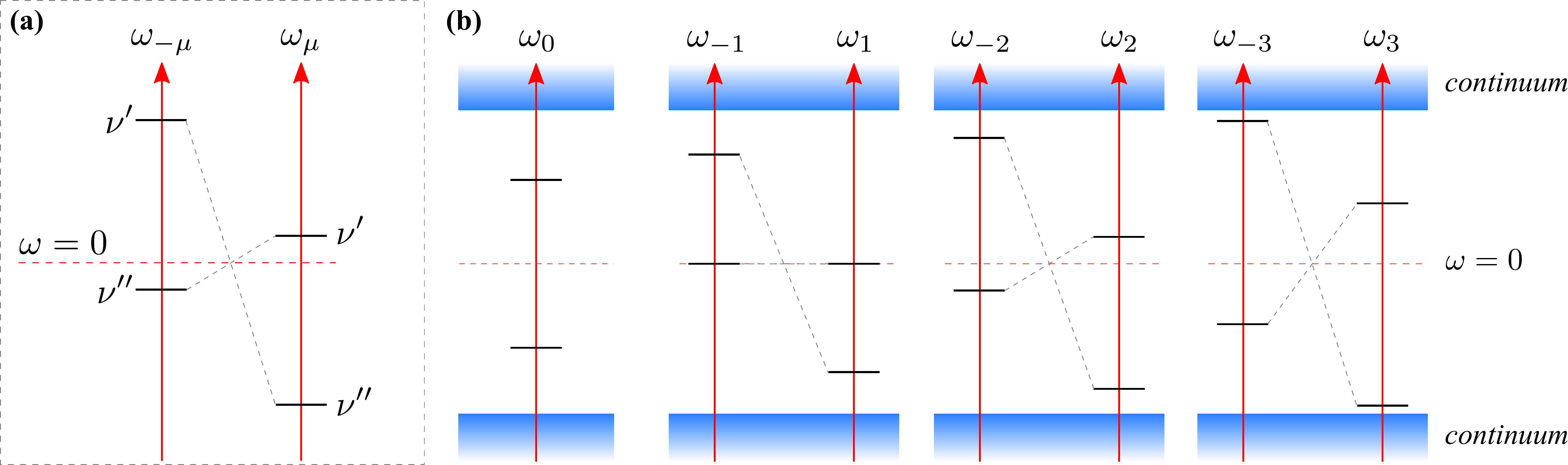}
	\caption{Schematics of the magnon spectrum. Panel (a) demonstrates the symmetry between branches $\mu$ and $-\mu$ (only one pair $\{\nu',\nu''\}$ is shown). Panel (b) shows the structure of the localized mode spectrum for $|\mu|\le3$. }\label{fig:specrum-scheme}
\end{figure*}

Equations \eqref{eq:EVP-mu-nu} are invariant with respect to the simultaneous transformation $\omega_\mu\to-\omega_{-\mu}$, $f_\mu\to f_{-\mu}$, $g_\mu\to-g_{-\mu}$, and  $\mu\to-\mu$. It means that the $\nu$-spectra can be split into the pairs $\bar{\nu}=\{\nu',\nu''\}$, which have the properties $\omega_{\mu,\nu'}=-\omega_{-\mu,\nu''}$,  $f_{\mu,\nu'}=f_{-\mu,\nu''}$, and $g_{\mu,\nu'}=-g_{-\mu,\nu''}$. This symmetry is schematically shown in the Fig.~\ref{fig:specrum-scheme}(a).
It can be utilized in order to present the general solution in a form, which coincides with \eqref{eq:dev-gen} but: (i) summation over $\nu$ is replaced by summation over pairs $\bar{\nu}=\{\nu',\nu''\}$ and for each pair the only one index $\nu\in\{\nu',\nu''\}$ is chosen, which corresponds to a \emph{certain sign of the frequency} $\omega_{\mu,\nu}$;
 (ii) $\bar{A}_{\mu,\nu}\to A_{\mu\nu}=\bar{A}_{\mu,\nu'}+\bar{A}_{-\mu,\nu''}$ and $\bar{B}_{\mu,\nu}\to B_{\mu\nu}=\bar{B}_{\mu,\nu'}-\bar{B}_{-\mu,\nu''}$.

Analyzing numerically the spectrum of the EVP \eqref{eq:EVP-mu-nu} we found a number of localized modes with $|\omega_{\mu,\nu}|<1$. Remarkably for each given $\mu$ there is no more than one pair $\bar{\nu}=\{\nu',\nu''\}$, which corresponds to the localized modes, see Fig.~\ref{fig:specrum-scheme}(b). 
This fact and the above described possibility to fix the sign of the frequency, enables one to omit index $\nu$ when classifying the localized modes. Thus, in the following we consider that for each given $\mu$ there is no more than one localized mode with eigenfunctions $f_\mu(\rho)$, $g_\mu(\rho)$, and eigenfrequency $\omega_\mu\ge0$ (the equality takes place for $\mu=1$ only). Several examples of eigenfunctions are shown in the Fig.~\ref{fig:eigenfunctions}.

The eigenfunctions have a number of useful properties. First of all, for the case of the localized modes, the described above symmetry between branches $\mu$ and $-\mu$ enable one to write the orthogonality condition \eqref{eq:orthog} in form \eqref{eq:norm}.

The next important property is integral relations between the eigenfrequencies and the eigenfunctions:
\begin{equation}\label{eq:omegas}
\omega_\mu=\frac{\varepsilon_\mu^f}{\alpha_\mu}=\frac{\varepsilon_\mu^g}{\alpha_\mu}=\frac{\kappa_\mu^f}{\beta_\mu}=\frac{\kappa_\mu^g}{\beta_\mu}.
\end{equation}
Here the condition
\begin{equation}\label{eq:eps-fg}
\varepsilon_\mu^f=\varepsilon_\mu^g,
\end{equation}
where
\begin{equation}
\begin{split}
&\varepsilon_\mu^f=\int\limits_0^\infty\rho\left[f_\mu'^2+\left(\frac{\mu^2}{\rho^2}+U_1\right)f_\mu^2+\mu Wf_\mu g_\mu\right]\mathrm{d}\rho,\\
&\varepsilon_\mu^g=\int\limits_0^\infty\rho\left[g_\mu'^2+\left(\frac{\mu^2}{\rho^2}+U_2\right)g_\mu^2+\mu Wf_\mu g_\mu\right]\mathrm{d}\rho,
\end{split}
\end{equation}
means that the energy is equally distributed between $f$ and $g$ components of the mode. And the condition
\begin{equation}\label{eq:kappa-fg}
\kappa_\mu^f=\kappa_\mu^g,
\end{equation}
where
\begin{equation}
\begin{split}
&\kappa_\mu^f=\int\limits_0^\infty\rho\left[f_\mu'f'_{-\mu}+\left(\frac{\mu^2}{\rho^2}+U_1\right)f_\mu f_{-\mu}+\mu Wf_{-\mu}g_\mu\right]\mathrm{d}\rho,\\
&\kappa_\mu^g=\int\limits_0^\infty\rho\left[g_\mu'g'_{-\mu}+\left(\frac{\mu^2}{\rho^2}+U_2\right)g_\mu g_{-\mu}+\mu Wf_\mu g_{-\mu}\right]\mathrm{d}\rho,
\end{split}
\end{equation}
results in absence of coupling between modes $|\mu\rangle$ and $|-\mu\rangle$. Here
\begin{equation}\label{eq:alpha-beta}
\alpha_\mu=\int\limits_0^\infty\rho f_\mu g_\mu\text{d}\rho,\quad \beta_\mu=\int\limits_0^\infty\rho f_\mu g_{-\mu}\text{d}\rho.
\end{equation}
Note that $\beta_\mu=\beta_{-\mu}$, while $\alpha_\mu\ne\alpha_{-\mu}$.
 For case of translational mode one has $\alpha_1=2$ and also
\begin{equation}\label{eq:transl}
\varepsilon_{1}^f=\varepsilon_{1}^g=\kappa_{1}^f=\kappa_{1}^g=0,
\end{equation} 
which results in zero eigenfrequency $\omega_1=0$.

All properties \eqref{eq:omegas}, \eqref{eq:eps-fg}, \eqref{eq:kappa-fg}, \eqref{eq:transl} directly follow from Eq.~\eqref{eq:EVP}, which have the explicit form
\begin{subequations}\label{eq:fg}
\begin{align}
\label{eq:f}&-\Delta_\rho f_\mu+\left(\frac{\mu^2}{\rho^2}+U_1\right)f_\mu+\mu Wg_\mu = \omega_\mu g_\mu,\\
\label{eq:g}&-\Delta_\rho g_\mu+\left(\frac{\mu^2}{\rho^2}+U_2\right)g_\mu+\mu Wf_\mu = \omega_\mu f_\mu.
\end{align}
\end{subequations}
Using the integration by parts one can present the quantity $\varepsilon_{\mu}^f$ in the form 
\begin{equation}\label{eq:eps-f}
\varepsilon_\mu^f=\int\limits_0^\infty\rho\left[-\Delta_\rho f_\mu+\left(\frac{\mu^2}{\rho^2}+U_1\right)f_\mu+\mu W g_\mu\right]f_\mu\mathrm{d}\rho.
\end{equation}
Applying now \eqref{eq:f} one obtains $\varepsilon_\mu^f=\omega_\mu\alpha_\mu$. In the similar way, using \eqref{eq:g} one obtains $\varepsilon_\mu^g=\omega_\mu\alpha_\mu$. The property \eqref{eq:kappa-fg} can be proved analogously by applying the same procedure for the quantities $\kappa_\mu^f$ and $\kappa_\mu^g$. The property \eqref{eq:transl} can be checked straightforwardly by substituting into \eqref{eq:fg} the explicit form of the eigenfunctions $f_1=-\Theta'$ and $g_1=\sin\Theta/\rho$.
All properties \eqref{eq:norm}, \eqref{eq:eps-fg}, \eqref{eq:kappa-fg}, \eqref{eq:omegas}, \eqref{eq:transl} have been verified numerically.


In order to estimate the asymptotic behavior of the eigenfrequencies for the case $R_s\gg1$ we apply the variational approach~\cite{Sheka01} with the following trial functions
\begin{equation}\label{eq:trial}
f_\mu=\frac{a_\mu}{\cosh\frac{\rho-R_s}{|\delta|}},\qquad g_\mu=\frac{b_\mu}{\rho\cosh\frac{\rho-R_s}{|\delta|}}.
\end{equation}
Substituting \eqref{eq:trial} into \eqref{eq:eps-fg} one obtains 
\begin{equation}\label{eq:eps-trial}
\begin{split}
&\varepsilon_\mu^f\approx\frac{2}{R_s}\left(\mu^2a_\mu^2-\mu a_\mu b_\mu\right)+\mathcal{C}_\mu^f\frac{a_\mu^2}{R_s^3}+\mu \mathcal{C}\frac{a_\mu b_\mu}{R_s^3},\\
&\varepsilon_\mu^g\approx\frac{2}{R_s}\left(b_\mu^2-\mu a_\mu b_\mu\right)+\mathcal{C}_\mu^g\frac{b_\mu^2}{R_s^3}+\mu \mathcal{C}\frac{a_\mu b_\mu}{R_s^3}.
\end{split}
\end{equation}
Generally $\mathcal{C}_\mu^f\ne\mathcal{C}_\mu^g$. This means that for the used ansatz \eqref{eq:trial} the conditions $\varepsilon_\mu^f=\varepsilon_\mu^g$ and $\varepsilon_1^f=\varepsilon_1^g=0$ can be satisfied simultaneously only with accuracy $\mathcal{O}(R_s^{-3})$. In this case $b_\mu=|\mu|a_\mu$. Using now the expressions for eigenfrequencies \eqref{eq:omegas} and taking into account that $\alpha_{\mu}\approx 2\delta a_\mu b_\mu$, one obtains $\omega_{\mu}\approx2|\mu|R_s^{-1}$ when $\mu<0$ and $\omega_{\mu}\approx c_\mu\,R_s^{-3}$ when $\mu>0$. The constant $c_\mu$ can not be determined in frames of the model \eqref{eq:trial}. Comparison with the numerical solution enables one to suppose that $c_\mu=\mu(\mu^2-1)/2$.

\section{Generalized Schr{\"o}dinger equation }\label{app:Schr}
Introducing function $\psi=\vartheta+i\varphi$ one can write \eqref{eq:dev} in form of so called generalized Schr{\"o}dinger equation \cite{Sheka04,Ivanov05b},
\begin{equation} \label{eq:Scroedinger}
-i\partial_\tau \psi = {\mathscr{H}} \psi + \mathcal{W} \psi^\star, \quad {\mathscr{H}} = \left(-i\vec{\nabla}-\vec{A}\right)^2 + \mathcal{U}
\end{equation}
with the potentials
\begin{equation} \label{eq:potentials}
\begin{split}
\vec{A}(\rho) &= A \vec{e}_\chi, \qquad A = -\frac{\cos\Theta}{\rho} + \frac{|d|}{2}\sin\Theta,\\
\mathcal{U}(\rho) &= \frac{U_1+U_2}{2} - A^2,\\
\mathcal{W}(\rho) &=\frac{U_1 - U_2}{2}.
\end{split}
\end{equation}
The vector function $\vec{A}$ acts as a vector potential of effective magnetic field with the flux density
\begin{equation} \label{eq:flux-density}
\begin{split}
\vec{B} = \vec{\nabla} \times \vec{A} = B \vec{\hat{z}},\quad B = \frac{\left(-\cos\Theta + \frac{|d|}{2}\rho\sin\Theta\right)'}{\rho} .
\end{split}
\end{equation}
The first term in \eqref{eq:flux-density} is the gyrocoupling density: finally the total flux is determined by $\pi_2$ topology of the skyrmion, cf. Ref.~\onlinecite{Sheka04}
\begin{equation} \label{eq:flux}
\Phi = \int B \,\mathrm{d}^2x = -4\pi \mathcal{Q}.
\end{equation}


%

%

\end{document}